%% file: main.tex
\documentclass[sigconf,natbib=false, authorversion]{acmart}

\usepackage{array,multirow,graphicx}
\usepackage{float}
\usepackage{multicol}
\usepackage{nicefrac} 
\usepackage{wasysym} 

\graphicspath{{img}}

\makeatletter
\newcommand\footnoteref[1]{\protected@xdef\@thefnmark{\ref{#1}}\@footnotemark}

\gdef\@copyrightpermission{
   \begin{minipage}{0.3\columnwidth}
     \href{https://creativecommons.org/licenses/by-sa/4.0/}{\includegraphics[width=0.90\textwidth]{img/4ACM-CC-by-sa-88x31.eps}}
   \end{minipage}\hfill
   \begin{minipage}{0.7\columnwidth}
     \href{https://creativecommons.org/licenses/by-sa/4.0/}{This work is licensed under a Creative Commons Attribution-ShareAlike International 4.0 License.}
   \end{minipage}
   \vspace{5pt}
}

\makeatother

\AtBeginDocument{%
  }

\copyrightyear{2024}
\acmYear{2024}
\setcopyright{rightsretained}
\acmConference[SIGIR '24] {Proceedings of the 47th International ACM SIGIR Conference on Research and Development in Information Retrieval}{July 14--18, 2024}{Washington, DC, USA.}
\acmBooktitle{Proceedings of the 47th International ACM SIGIR Conference on Research and Development in Information Retrieval (SIGIR '24), July 14--18, 2024, Washington, DC, USA}
\acmISBN{979-8-4007-0431-4/24/07}
\acmDOI{10.1145/3626772.3657832}

\RequirePackage[
  datamodel=acmdatamodel,
  style=acmnumeric,
  sortcites=true 
  ]{biblatex}

\DeclareSourcemap{
  \maps{
    \map{
      \step[fieldsource=doi,
        match=\regexp{\{\\_\}},
        replace=\regexp{_}]
      \step[fieldsource=author,
        match=\regexp{\'\{\\i\}},
        replace=\regexp{í}]
    }
  }
}

\begin{document}
\title{Can We Trust Recommender System Fairness Evaluation? The Role of Fairness and Relevance}


\author{Theresia Veronika Rampisela}
\email{thra@di.ku.dk}
\orcid{0000-0003-1233-7690}
\affiliation{%
  \institution{University of Copenhagen}
  \city{Copenhagen}
  \country{Denmark}}

\author{Tuukka Ruotsalo}
\email{tr@di.ku.dk}
\orcid{0000-0002-2203-4928}
\affiliation{%
 \institution{University of Copenhagen}
 \city{Copenhagen}
 \country{Denmark}
}
\affiliation{%
 \institution{LUT University}
 \city{Lahti}
 \country{Finland}
}

\author{Maria Maistro}
\email{mm@di.ku.dk}
\orcid{0000-0002-7001-4817}
\affiliation{%
 \institution{University of Copenhagen}
 \city{Copenhagen}
 \country{Denmark}
}

\author{Christina Lioma}
\email{c.lioma@di.ku.dk}
\orcid{0000-0003-2600-2701}
\affiliation{%
 \institution{University of Copenhagen}
 \city{Copenhagen}
 \country{Denmark}
}


\newcommand{\rs}{RSs}

\newcommand{\up}{$\uparrow$}
\newcommand{\down}{$\downarrow$}

\begin{abstract}{
Relevance and fairness are two major objectives of recommender systems (RSs). 
Recent work proposes measures of RS fairness that are either independent from relevance (fairness-only) or conditioned on relevance (joint measures). 
While fairness-only measures have been studied extensively, we look into whether joint measures can be trusted. 
We collect all joint evaluation measures of RS relevance and fairness, and ask: How much do they agree with each other? To what extent do they agree with relevance/fairness measures? How sensitive are they to changes in rank position, or to increasingly fair and relevant recommendations? 
We empirically study for the first time the behaviour of these measures across 4 real-world datasets and 4 recommenders. We find that most of these measures: 
i) correlate weakly with one another and even contradict each other at times; 
ii) are less sensitive to rank position changes than relevance- and fairness-only measures, meaning that they are less granular than traditional RS measures; and 
iii) tend to compress scores at the low end of their range, meaning that they are not very expressive. 
We counter the above limitations with a set of guidelines on the appropriate usage of such measures, i.e., they should be used with caution due to their tendency to contradict each other and of having a very small empirical range. 
}
\end{abstract}



\maketitle

\section{Introduction}
Recent increased focus on fairness in recommender systems (RSs) has led to studies on how to evaluate different notions of fairness in RS. A recent survey \cite{Wang2022} shows that prior work on fairness evaluation in RS mainly focuses on group fairness (e.g., \cite{Raj2022MeasuringResults,Amigo2023ASystems,Zehlike2022FairnessSystems}), but less so on individual fairness. 
Individual fairness is commonly understood as treating similar individuals similarly \cite{Dwork2012FairnessAwareness}. 
Unlike group fairness evaluation, evaluating individual fairness does not require information on sensitive attributes (e.g., gender, age) to identify protected groups \cite{Lazovich2022MeasuringMetrics}. Such information is often unavailable due to privacy and legal issues. Further, intersectionality between different group characteristics complicates group fairness \cite{Crenshaw1991MappingColor,Ekstrand2022FairnessSystems}. Individual fairness is known to lead into group fairness, but not vice versa \cite{Bower2021IndividuallyRanking}. 
Overall, individual fairness gives a broader view by assessing distribution across all individuals in the population \cite{Lazovich2022MeasuringMetrics}. For all these reasons, we focus on individual fairness, particularly individual item fairness, which is typically broadly defined w.r.t.~exposure received by items, i.e., how uniform the exposure distribution between items is \cite{Rampisela2023EvaluationStudy}. Yet, fairness beyond exposure also matters, i.e., the exposure should be proportional to item relevance \cite{Patro2022FairDirections,Smith2023ScopingPerspective,Biega2018EquityRankings}. 

Individual item fairness is measured by measures that (i) are detached from relevance (fairness-only measures, defined by exposure); or (ii) are conditioned on relevance (joint measures considering exposure w.r.t.~relevance). Measures of type (i) have been extensively analysed \cite{Rampisela2023EvaluationStudy}, but to our knowledge, this is not the case for measures of type (ii). The growing number of measures of type (ii) necessitates a thorough look into their usage in RS evaluation. 

We present a comprehensive study into the empirical properties of all joint measures of individual item fairness and relevance, motivated by the question of how much can we practically trust these measures, particularly: 
\textbf{RQ1.} To what extent do the joint measures agree with existing relevance- and fairness-only measures? 
\textbf{RQ2.} To what extent do the joint measures agree with each other?
\textbf{RQ3.} How sensitive are the joint measures across decreasing rank positions? and 
\textbf{RQ4.} How sensitive are the joint measures given increasingly fair and relevant recommendations?

We identify some alarming limitations in the measures, and we reflect on their best usage in practice. This is the first in-depth study on individual item fairness measures that consider relevance in RS.

\section{Individual item fairness \& relevance}
\label{s:priorwork}

We present the notation ($\S$\ref{ss:notation}) and all existing joint evaluation measures of individual item fairness and relevance ($\S$\ref{ss:fairrel}). 

\subsection{Notation and definitions}
\label{ss:notation}

Given a set of $n$ items, $I = \{i_1, i_2, \dots, i_n\}$, and a set of $m$ users, $U=\{u_1, u_2, \dots, u_m\}$, an ordered 
list of the $n$ items is created for each $u \in U$. This list is created in each recommendation round $w$, where $w \in \{1, 2, \dots, W\}$; a round means an occurrence in which a user receives a list of recommendations. 
If an item $i$ is \emph{relevant} to user $u$, we write $r_{u,i}=1$, otherwise $r_{u,i}=0$. Relevance can also be denoted as real values, $r_{u,i}\in[0,1]$. The list of user $u$'s top $k$ recommended items in round $w$ is $L_{u, w}$ and the rank position of item $i$ in user $u$'s recommendation list is $z(u,i,w)$. 
For cases with only one recommendation round, user $u$'s list of top $k$ recommended items is $L_u$ and the rank position of item $i$ for user $u$ is $z(u,i)$. 

While several different definitions of fairness exist, the definitions commonly used in prior work on individual item fairness are closely linked to item exposure 
\cite{Rampisela2023EvaluationStudy,Amigo2023ASystems,Zehlike2022FairnessSystems}. An item is \emph{exposed} when it is recommended at the top $k$ to a user. 
The probability of a user seeing an item exposed to them can be modelled using various \emph{examination functions}, $e(\cdot)$. Examination functions typically assume that the viewing probability depends only on the position $z(u,i,w)$ or $z(u,i)$. This is a common choice across all measures in $\S$\ref{ss:fairrel}. 

The examination functions used in this work are shown in Tab.~\ref{tab:exp-weigh}: the linear examination function, $e_{\text{li}}$ and its min-max normalised version $\tilde{e}_{\text{li}}$ apply a linear discount to each rank position up to $k$ \cite{Borges2019EnhancingAutoencoders}.
Meanwhile, discounts based on Discounted Cumulative Gain (DCG)~\cite{Jarvelin2002CumulatedTechniques} and Rank-Biased Precision (RBP)~\cite{Moffat2008Rank-biasedEffectiveness} are used in $e_{\text{DCG}}$  \cite{Singh2019PolicyRanking,Oosterhuis2021ComputationallyFairness} 
and $e_{\text{RBP}}$ \cite{Wu2022JointRecommendation,Jeunen2021Top-KExposure} respectively. The parameter $\gamma$ in $e_{\text{RBP}}$ is the user's patience, i.e., the probability of the user examining the next ranked item. The user patience parameter is commonly set at 
e.g., $\gamma \in \{0.8, 0.9\}$  \cite{Wu2022JointRecommendation, Jeunen2021Top-KExposure}.
In the inverse examination function $e_{\text{inv}}$, the inverse of the rank position is used as a discount factor \cite{Saito2022FairRanking}. Overall, we use three types of examination functions (linear, discounted, and inverse), which assume that item exposure diminishes with decreasing ranking either linearly, or with an increasing penalty that is either proportional to the rank decrease or the inverse of the rank. Generally, the most punishing is the inverse function, and the least punishing is the DCG-based discount function.

\begin{table}[b]
\caption{Examination functions used in this work. $\tilde{e}$ is the min-max normalised examination function.}
\label{tab:exp-weigh}
\resizebox{\columnwidth}{!}{
\begin{tabular}{llll}
\toprule
         & Equation & Measure & Reference                        \\ 
         \midrule
linear   & \parbox[t]{5cm}{$e_{\text{li}}(u, i, w) = k+1-z(u,i,w)$ \\
    $\tilde{e}_{\text{li}}(u, i, w) = \frac{ e_{\text{li}}(u, i, w)-1}{k-1} = \frac{k-z(u,i,w)}{k-1}$}                                                   & IAA   &  \cite{Borges2019EnhancingAutoencoders}                        \\
DCG      & $e_{\text{DCG}}(u, i, w) = 1/\log_2 (z(u,i,w)+1)$         & IFD & \cite{Singh2019PolicyRanking, Oosterhuis2021ComputationallyFairness}   \\
RBP      & $e_{\text{RBP}}(u,i,w) = \gamma^{z(u,i,w)-1}$         & HD, II-F, AI-F   & \cite{Wu2022JointRecommendation, Jeunen2021Top-KExposure}                 \\
inverse  & $e_{\text{inv}}(u,i) = 1/z(u,i)$                                & MME, IBO/IWO & \cite{Saito2022FairRanking}        \\
\bottomrule
\end{tabular}
}
\end{table}

\subsection{Joint measures of fairness and relevance}
\label{ss:fairrel}

We present measures that evaluate fairness considering relevance (\textsc{Fair+Rel} or joint measures henceforth). 
To our knowledge, we include all \textsc{Fair+Rel} measures for RSs published up to October 2023. Each measure uses an exposure function, which is linked to the fairness of item distribution in the recommendation and, therefore, measures item fairness jointly with relevance. 
We use \up~for measures where the higher the score, the fairest the recommendation, and vice versa for \down. 
All measures--except HD--are defined 
for multiple recommendation rounds or stochastic rankings, where a distribution over rankings is considered \cite{Biega2018EquityRankings}.

\subsubsection{Inequity of Amortized Attention (IAA) \cite{Biega2018EquityRankings}}
\label{ss:iaa-ori} 
\down IAA\footnote{This measure is called IAA in \cite{Raj2022MeasuringResults} and L1-norm in \cite{Wang2022}.} measures fairness as the aggregated difference between item exposure and its relevance in a series of rankings that have been generated by a stochastic process \cite{Biega2018EquityRankings}. 
The intuition behind IAA is that for a sequence of rankings to be fair, 
items should be allocated exposure proportional to their relevance to the user. 
The item position is a proxy of its exposure level. 
IAA was modified in~\cite{Borges2019EnhancingAutoencoders} to account for multiple recommendation rounds (stochastic rankings): 
\begin{equation}
    \label{eq:iaa-ori-rel}
    \text{IAA} = \frac{1}{m}\sum\limits_{u \in U}\text{IAA}(u) 
    \end{equation}%
    \begin{equation}
    \label{eq:iaa-ori-u}
    \text{IAA}(u) = \frac{1}{n}\frac{1}{W}
        \sum_{i \in I}
        \left|
        \sum_{w=1}^W 1_{L_{u,w}}(i) \cdot \tilde{e_{.}}(u,i,w)
        -  
        \tilde{r}(u,i,w) 
        \right|
\end{equation}

In~\cite{Borges2019EnhancingAutoencoders}, 
 $\tilde{e}_{(\cdot)}(u,i,w)$ is the min-max normalised linear examination function $\tilde{e}_{\text{li}}(u, i, w)$ (see Tab.~\ref{tab:exp-weigh}) 
and $\tilde{r}(u,i,w) \in [0,1]$ is the min-max normalised relevance 
value of item $i$ for user $u$ in round $w$, $r_{u,i,w}$.\footnote{Note that the normalised exposure value for a recommended item at $k$ is zero.}
Both the min and max relevance values are taken from the values associated with all items for each user per round, i.e., $\min_{i \in I} r_{u,i,w}$. 
This value is the aggregated relevance over all items for user $u$ in round $w$; the higher the relevance, the closer to 1. The higher the relevance value differs from item exposure, the more unfair. The range of IAA is $[0,1]$. 

\subsubsection{Individual Fairness Disparity (IFD) \cite{Singh2019PolicyRanking,Oosterhuis2021ComputationallyFairness}} \down IFD is the average pairwise difference of the combined value of item exposure and item merit. Merit is defined as a function of relevance.\footnote{We use the item relevance value as the item merit, as per \cite{Singh2019PolicyRanking}.} 
Similar to IAA, IFD follows the principle of allocating exposure to an item based on its relevance. 
While IAA computes the difference between the exposure and relevance of each item, IFD computes the disparity of exposure allocation between item pairs. 
Based on how exposure and merit are combined, two variations of IFD exist: IFD$_{\div}$, where item exposure is divided by item relevance \cite{Singh2019PolicyRanking}, 
and IFD$_{\times}$, where the division is replaced by multiplication \cite{Oosterhuis2021ComputationallyFairness}. The term IFD$_{(\cdot)}$ or IFD refers to the measure in general. 
The two versions slightly differ in the pairwise difference computation, the formation of set of item pairs, and the exposure weighting scheme.\footnote{Exposure is weighed proportional to $e_{\text{DCG}}$ in \cite{Singh2019PolicyRanking}; to simplify, we use $e_{\text{DCG}}$ directly.} 
Both IFD versions have been used to measure fairness in ranking \cite{Singh2019PolicyRanking, Oosterhuis2021ComputationallyFairness, Yang2023FARA:Optimization, Yang2023Marginal-Certainty-AwareAlgorithm}. 
\begin{equation}
\label{eq:IFD}
\text{IFD}_{(\cdot)} = \frac{1}{m} \sum\limits_{u \in U}\text{IFD}_{(\cdot)}(u) 
\end{equation}
\begin{equation}
\label{eq:IFD-u-div}
\text{IFD}_{\div}(u) = 
\frac{1}{|H_u|} \sum_{(i,i')\in H_u} \max{ \left\{0, 
J_{\div}(u,i) - J_{\div}(u,i')
\right\}
} 
\end{equation}
\begin{equation}
\label{eq:IFD-u-mult}
\text{IFD}_{\times}(u) = 
\frac{1}{n(n-1)} \sum_{i \in I} \sum_{i' \in I\setminus{i}}
\left[
    J_{\times}(u,i) - J_{\times}(u,i')
\right]^2
\end{equation}
\begin{equation}
    J_{\div}(u,i) = \frac{1}{W} \sum\limits_{w=1}^{W} e_{\text{DCG}}(u,i,w)/r_{u,i,w}
\end{equation}
\begin{equation}\label{eq:ifd_j_x}
    J_{\times}(u,i) = \frac{1}{W} \sum\limits_{w=1}^{W} 
    r_{u,i,w} \cdot 1_{L_{u,w}}(i) \cdot e_{\text{DCG}}(u,i,w) 
\end{equation}
 $J_{(\cdot)}(u,i)$ is the function combining the expected exposure and relevance of item $i$ for user $u$ and $H_u = \{(i,i') \in I\ |\ r_{u,i} \geq r_{u,i'} > 0\}$.
The range of IFD$_{\div}$ is $[0,\infty)$ and it is 0 only when the exposure received by each relevant item is exactly proportional to its relevance \cite{Singh2019PolicyRanking}. The range of IFD$_{\times}$ is $[0,\infty)$ based on empirical results \cite{Yang2023FARA:Optimization}. 

\subsubsection{Hellinger Distance (HD) \cite{Jeunen2021Top-KExposure}}  
\down HD has been used as a measure of individual item fairness in top $k$ contextual bandits, by quantifying the difference between the relevance- and click-distributions of the top $k$ items sorted according to (ground truth) relevance \cite{Jeunen2021Top-KExposure}. The click probability is based on user patience, system-allocated item exposure, and item relevance.
A recommendation is fair based on HD when the click probability of an item is proportional to the relevance probability of that item. To compute the relevance and click distributions, a list of top $k$ items is created for each user by sorting items based on their (ground truth) relevance; this list is the reference list used in the next step. Another list of items is created based on system prediction and used to get the click probability. For each item in the reference list, we compute its click probability based on its order in the second list. Next, the relevance probabilities of items at the same position in the reference list are aggregated across users and similarly for the click probabilities. For each rank position, two aggregated values are obtained: relevance and click. The aggregated values are the inputs to the distance metric (Eq.~\eqref{eq:HD}).
\begin{equation}\label{eq:HD}
    \text{HD} = \frac{1}{\sqrt{2}} \sqrt{
     \sum_{p=1}^k \left(\sqrt{q_p'} - \sqrt{c_p'}\right)^2}
\end{equation}
\begin{equation}
    q_p' = \frac{1}{m} \sum_{u \in U} \sum_{i \in I} 
    \delta\left(z^*(u,i) = p\right) \cdot r'_{u,i}
\end{equation}
\begin{equation}
     c_p' = \frac{1}{m} \sum_{u \in U} \frac{c^*_{u,p}}{\sum_{\ell=1}^{k}c^*_{u,\ell}}
\end{equation}
\begin{equation}
    c^*_{u,p} = \sum_{i \in I} 
    \delta\left(z^*(u,i) = p\right) \cdot c^{full}_{u,i}
\end{equation}
\begin{equation}
    c^{full}_{u,i} = c'_{u,p}\, \text{if } \exists p : z(u,i)=p\, \text{, otherwise } 0
\end{equation}
\begin{equation}
    c_{u,p} = \sum_{i \in L_u} 
    \delta\left(z(u,i) = p\right) \cdot r_{u,i} \cdot \gamma \ e_{\text{RBP}}(u,i) \cdot s_{u,p}
\end{equation}
\begin{equation}
    s_{u,p} = \prod_{1 \le j < p} 1 - \sum_{i \in L_u} \delta(z(u,i) = j) \cdot r_{u,i} 
\end{equation}

\noindent where $q_p'$ and $c_p'$ are the normalised relevance and click probability of the item at position $j$ respectively, where click depends on both relevance and exposure. The position of item $i$ based on ground-truth relevance is $z^*(u,i)$. The click probability of user $u$ for item at position $p$, $c_{u,p}$ depends on $s_{u,p}$, the probability that items before position $p$ were irrelevant to the user, and the user patience $\gamma \ e_{\text{RBP}}(u,i)$. 
$r'_{u,i} = r_{u,i}/\sum_{i \in I} r_{u,i}$
is the user-wise normalised relevance value of item $i$ to user $u$, and 
$c'_{u,p} = c_{u,p}/\sum^{k}_{p=1} c_{u,p}$ 
is the user-wise normalised click probability. The value of $\delta(\cdot) = 1$ when the expression $\cdot$ is True and 0 otherwise. HD ranges between $[0,\infty)$. 

\subsubsection{Mean Max Envy (MME) \cite{Saito2022FairRanking}}\label{sss:mme} 
\down MME uses the concept of envy-freeness, where a recommendation is fair when each item is not disadvantaged by its own exposure allocation compared to being allocated the exposure of any other item. In other words, MME computes unfairness as the disadvantage suffered by the item, if the exposure allocation of an item is swapped with another item. The disadvantage is computed based on an impact score that uses exposure and relevance: given full recommendation lists (size $n$) across all users, 
we swap each item $i$ with another item $i'$ and compute the impact score before and after the swap for all rank positions and users. If the score of the item $i$ before the swap is greater or equal to its score after the swap, we have envy-freeness for item $i$ w.r.t.~item $i'$.
MME thus computes the average maximum difference of impact 
imposed if item $i$ is replaced with another item $i'$. 
E.g., let $L_{u_1} = [i_1, i_2],\  L_{u_2} = [i_1, i_3]$ and let us swap item $i_3$ with $i_1$. Item $i_3$ will be exposed to both users at the top position, like $i_1$ did, and then impact is recomputed. 
MME is computed as follows:

\begin{equation}
    \label{eq:mme-ori}
    \text{MME} = \frac{1}{n} \sum_{i \in I} 
    \left\{
        \max_{i' \in I} Imp_i(i') - Imp_i(i)  
    \right\}
\end{equation}
\begin{equation}
        \label{eq:imp}
        Imp_i(i') = \sum\limits_{u \in U} \sum\limits_{p=1}^k r_{u,i} \cdot e_{\text{inv}}(u,i') \cdot X_{u,i',p} 
\end{equation}
\begin{equation}
     X_{u,i',p} = \frac{1}{W} \frac{1}{m} \sum\limits_{w=1}^W 1_{L_{u,w}}(i') \cdot \delta(z(u,i',w)=p)  
\end{equation}
where 
$Imp_i(i')$ is the impact when we allocate the exposure of item $i'$ to item $i$, 
$X_{u,i',p}$ is the probability that item $i'$ is recommended to user $u$ at position $p$ in $W$ rounds of recommendations, 
and $e_{\text{inv}}(u,i')$ is the exposure weight of item $i'$ to user $u$, based on the inverse examination function (see Tab.~\ref{tab:exp-weigh}). MME ranges within $[0,\infty)$. 

\subsubsection{Item Better-Off (IBO) \& Item Worse-Off (IWO) \cite{Saito2022FairRanking}} \up IBO and \down IWO use the principle of \textit{dominance over uniform ranking}, where fairness means each item 
has a better impact (as defined in MME) under the current ranking policy, than if it were under the uniform random ranking policy $\pi_{unif}$, which samples all possible permutations of items uniformly at random. 
IBO/IWO measures the percentage of items for which our current ranking policy increases/decreases impact by at least 10\% compared to $\pi_{unif}$\footnote{In \cite{Saito2022FairRanking}, 10\% is hard coded, but this can be a variable. We also use 10\%.}:
\begin{equation}
    \label{eq:ibo-our}
     \text{IBO} = \frac{100}{|I^-|} 
     \sum_{i \in I^-} 
   \delta\left(Imp_i(i) \geq 1.1 \cdot Imp_{i}^{unif}\right)
\end{equation}
\begin{equation}
    \label{eq:iwo-our}
     \text{IWO} = 
     \frac{100}{|I^-|} 
     \sum_{i \in I^-} 
    \delta\left(Imp_i(i) \leq 0.9 \cdot Imp_{i}^{unif}\right) 
\end{equation}
\begin{equation}
\label{eq:imp-unif}
    Imp^{unif}_{i} =  \frac{1}{m}\frac{1}{n} \sum \limits_{p=1}^{k} \frac{1}{p} \cdot \sum \limits_{u \in U} r_{u,i} 
\end{equation}

\noindent where $I^{-}$ is the set of items with at least one user that finds the item relevant. This ensures that the set of items that cause $\delta(\cdot)=1$ in IBO is disjoint from that in IWO.\footnote{We exclude items with no relevant users, as 
for these items $Imp_i(i)=Imp^{unif}_i=0$, causing the same items being considered `better-off' and `worse-off' at the same time.} 
$Imp_i(i)$ is as per Eq.~\eqref{eq:imp} and $Imp^{unif}_i$ is the impact if item $i$ is exposed according to $\pi_{unif}$ using $e_{\text{inv}}(u,i)$ as examination function (see Tab.~\ref{tab:exp-weigh}). 
Note that the above definitions are modifications to the formulation of \cite{Saito2022FairRanking} to avoid computational issues that result in 
division by zero (undefinedness limitation \cite{Rampisela2023EvaluationStudy}).\footnote{We move the divisor $Imp^{unif}_i$ from the left-hand side to the right-hand side.} 
IBO/IWO ranges between $[0,100]$.

\subsubsection{Individual-user-to-individual-item fairness (II-F) \cite{Wu2022JointRecommendation}} 
\down II-F was first defined by \cite{Diaz2020EvaluatingExposure} to quantify unfairness as the disparity between system exposure and target exposure in individual queries and individual items. II-F was redefined by \cite{Wu2022JointRecommendation} for RSs as: 
\begin{equation}
    \label{eq:iif-ori}
     \text{II-F} = \frac{1}{m} \frac{1}{n} 
     \sum\limits_{u \in U} \sum\limits_{i \in I} \left(E_{u,i}^{ } - E_{u,i}^*\right)^2 
\end{equation}
\begin{equation}
    \label{eq:eui}
E_{u,i} = \frac{1}{W} \sum\limits_{w=1}^{W}1_{L_{u,w}}(i) \cdot e_{\text{RBP}}(u,i,w)
\end{equation}
\begin{equation}
\label{eq:eui-star}
E_{u,i}^{*} = \frac{r_{u,i}}{|R_u^*|} \cdot\frac{1-\gamma^{|R_u^*|}}{1-\gamma}  \, \text{if } |R_u^*|>0\, \text{, otherwise }0
\end{equation}
\noindent where $E_{u,i}$ is the expected exposure of $i$ to $u$ as per a stochastic ranking policy. 
$E_{u,i}^{*}$ is the expected exposure of $i$ to $u$ as per an ideal stochastic ranking policy, which assumes that relevant items get equal expected exposure~\cite{Diaz2020EvaluatingExposure}. 
Thus, the recommendation is fair based on II-F if the system exposure 
matches the exposure allocated to items under an ideal ranking policy.
The examination function based on RBP (see Tab.~\ref{tab:exp-weigh}) is used in $E_{u,i}$ and the equation of $E_{u,i}^{*}$ is derived based on the same examination function \cite{Wu2022JointRecommendation}. 
$|R_u^*|$ is the number of relevant items for user $u$. II-F ranges between $[0,1]$.

\subsubsection{All-users-to-individual-item fairness (AI-F) \cite{Wu2022JointRecommendation}} \down AI-F evaluates how much \rs~ under/overexpose an item to all users as the mean deviation of overall system exposure over target exposure:
\begin{equation}
    \label{eq:aif-ori}
     \text{AI-F} = 
        \frac{1}{n} \sum \limits_{i \in I}
            \left(
                \frac{1}{m}\sum \limits_{u \in U} E_{u,i}^{ }
                - \frac{1}{m}\sum \limits_{u \in U} E_{u,i}^{*}
            \right)^2
\end{equation}
where $E_{u,i}^{ }$, $E_{u,i}^{*}$ are as per Eq.~\eqref{eq:eui}--\eqref{eq:eui-star}. 
Similar to II-F, AI-F also quantifies fairness based on how close the system exposure is to the target exposure. In II-F, this disparity is computed individually between each user-item pair, while in AI-F item exposure is first aggregated across users prior to computing the difference in exposure. Due to this aggregation,  
AI-F would have a better fairness score than II-F when there is a greater number of unique items in the recommendation, as opposed to having the same few items exposed to all users. The range of AI-F is $[0,1]$. 

\section{Experimental setup}
\label{ss:setup}

We study the above \textsc{Fair+Rel} measures across different recommenders and datasets. 
Our general experimental setup follows, and we provide the description of the experiments in $\S$\ref{s:exp}.\footnote{Our code: \url{https://github.com/theresiavr/can-we-trust-recsys-fairness-evaluation}.}

\noindent \textbf{Datasets.} We use four real-world datasets of varying sizes and domains: 
Lastfm (music) \cite{Cantador20112nd2011}, 
Amazon Luxury Beauty, i.e., Amazon-lb (e-commerce) \cite{Ni2019JustifyingAspects},  
QK-video (videos) \cite{Yuan2022Tenrec:Systems}, and
ML-10M (movies) \cite{Harper2015TheContext}. QK-video is as provided by  \cite{Yuan2022Tenrec:Systems}, and the rest are as provided by \cite{Zhao2021RecBole:Algorithms}. For QK-video, we use only the `sharing' interactions.

\begin{table}[tb]
\caption{Statistics of 
the preprocessed datasets.}
\label{tab:dataset}

\resizebox{\columnwidth}{!}{
\begin{tabular}{lrrrr}
\toprule
\textbf{dataset} & \multicolumn{1}{l}{\textbf{\#users ($m$)}} & \multicolumn{1}{l}{\textbf{\#items ($n$)}} & \multicolumn{1}{l}{\textbf{\#interactions}} & \multicolumn{1}{l}{\textbf{sparsity (\%)}} \\ 
\midrule                                                           
Lastfm \cite{Cantador20112nd2011} & 1,859 & 2,823 & 71,355 & 98.64\% \\
Amazon-lb \cite{Ni2019JustifyingAspects} & 1,054 & 791 & 12,397 & 98.51\% \\
QK-video \cite{Yuan2022Tenrec:Systems} & 4,656 & 6,423 & 51,777 & 99.83\% \\ 
ML-10M \cite{Harper2015TheContext} & 49,378 & 9,821 & 5,362,685 & 98.89\% \\
\bottomrule
\end{tabular}
}
\end{table}

\noindent \textbf{Preprocessing.} We keep only users and items with at least 5 interactions (5-core filtering). When there are duplicate interactions, we keep the most recent one. Ratings equal/above 3 are converted to 1, and the rest are discarded for Amazon-lb and ML-10M, as their ratings range between $[1,5]$ and $[0.5, 5]$ respectively. No conversions are done for Lastfm and QK-video as they only have implicit feedback. Tab.~\ref{tab:dataset} presents statistics of the preprocessed datasets. 

\noindent \textbf{Data splits.} Global temporal splits \cite{Meng2020ExploringModels} with a ratio of 6:2:2 form the train/val/test sets from the preprocessed datasets for Amazon-lb and ML-10M. Global random splits with the same ratio are used for Lastfm and QK-video as they have no timestamps. 
Only users with at least 5 interactions in the train set are kept in all splits. 

\noindent \textbf{Recommenders.} We use four well-known top $k$ recommenders: 
item-based K-Nearest Neighbour (ItemKNN) \cite{Deshpande2004Item-basedAlgorithms}, Bayesian Personalised Ranking (BPR), \cite{RendleBPR:Feedback}, Variational Autoencoder with multinomial likelihood (MultiVAE) \cite{Liang2018VariationalFiltering}, and Neighbourhood-enriched Contrastive Learning (NCL) \cite{Lin2022ImprovingLearning}. 
We train BPR, MultiVAE, and NCL using RecBole \cite{Zhao2021RecBole:Algorithms} for 300 epochs with early stopping. The configuration with the best NDCG@10 during validation is taken as the final model.\footnote{The hyperparameter search space and best values are in the code repository.} 
During testing, all unobserved items are selected as candidates for recommendation and each user's train/val items are excluded from their own recommendations. 

\noindent \textbf{Fair re-ranker.} 
As the models are not directly optimised for fairness, we use a re-ranker to obtain fairer recommendations. 
The top $k'$ items are re-ranked to provide exposure to items that were outside the top $k$, where $k'$ is ideally larger than the cut-off $k=10$. In RS datasets, normally there are very few relevant items per user, so $k'$ should not be too big (e.g., 100). We choose $k'=25$ for all datasets and models. The re-ranking is done per user with COMBMNZ (CM) \cite{Lee1997AnalysesCombination} as a robust rank fusion method.\footnote{Other re-rankers exist but do not suit  
our setup, e.g., \cite{Wang2022ProvidingSystems} requires computing item similarity, but true similarity is challenging to obtain \cite{Dwork2012FairnessAwareness,Tsepenekas2023ComparingDistributions}.} 
CM fuses two lists of scores, one based on relevance and one based on fairness, to create a new ranking for each user. 
The relevance-based score is the min-max normalised predicted relevance score. The fairness-based score is first obtained from the coverage score of each top $k'$ items based on their appearance in the top $k$. Then, we compute 1 minus the normalised coverage to allocate higher score for items with lower exposure, thus increasing fairness. 
The combined scores are sorted to generate the final fused ranking of relevance and fairness. 

\noindent\textbf{Measures}. 
Recommendation models are evaluated using all the joint measures of relevance and fairness (\textsc{Fair+Rel}) presented in $\S$\ref{ss:fairrel}.\footnote{For IAA, the ground truth relevance is used to compute the relevance score. For HD, $\gamma=0.9$ as per 
\cite{Jeunen2021Top-KExposure}. For II-F and AI-F, $\gamma=0.8$  as per 
\cite{Wu2022JointRecommendation}. Note that 
IBO/IWO are normalised to [0,1] for consistency with the other measures.} 
As comparison to the joint measures, we evaluate relevance only (\textsc{Rel}) with: Hit Rate (HR), MRR, Precision (P), Recall (R), MAP, and NDCG. We also evaluate fairness only (\textsc{Fair}) with:\footnote{We use the modified versions of these measures as per \cite{Rampisela2023EvaluationStudy}.} 
Jain Index (Jain) \cite{jain1984quantitative,Zhu2020FARM:APPs}, Qualification Fairness (QF) \cite{Zhu2020FARM:APPs}, Entropy (Ent) \cite{Patro2020FairRec:Platforms,Shannon1948ACommunication},
Fraction of Satisfied Items (FSat) \cite{Patro2020FairRec:Platforms}, and Gini Index (Gini) \cite{Gini1912VariabilitaMutabilita,Mansoury2020FairMatch:Systems}. Unless otherwise stated, all measures are computed at $k=10$.

\section{Empirical analysis}
\label{s:exp}
We present the evaluation results of all \textsc{Fair+Rel}, \textsc{Rel}, and \textsc{Fair} measures, in $\S$\ref{ss:performance}. 
We study their correlation in $\S$\ref{ss:corr}, their sensitivity across different top-$k$ positions in $\S$\ref{ss:sliding} and across increasing levels of relevance and fairness in $\S$\ref{ss:insert}.

\subsection{Evaluation results of all measures}
\label{ss:performance}

\input{tab/base_table}

Tab.~\ref{tab:base-rerank-all} shows the scores of all \textsc{Fair+Rel}, \textsc{Rel}, and \textsc{Fair} measures, per dataset and recommender/re-ranking. $\uparrow$ means the higher the score, the better, and vice versa for $\downarrow$. Overall we observe the following. 

\noindent \textbf{Best model agreement.} We aim to study whether the measures agree on the same best model. We note two main trends. First, for all datasets, the best model based on \textsc{Rel} measures is always different from the one based on \textsc{Fair} measures, except for QF in Amazon-lb. This means that the fairest model is not necessarily the best in terms of relevance. 
Second, while all \textsc{Rel} measures agree on the same best model per dataset (except MRR and MAP for Amazon-lb) and all the \textsc{Fair} measures always agree on the same best model (except QF), the \textsc{Fair+Rel} measures disagree on the best model. Occasionally, some \textsc{Fair+Rel} measures agree with another more often (e.g., IBO with IWO, or IAA with HD and II-F, or MME with AI-F and sometimes IFD), but there is no overall consistency. The agreement between some joint measures may be due to their similar formulations: both IBO/IWO are the fractions of items with an impact score greater/lower than a threshold; MME/AI-F aggregate exposure across users prior to computing the exposure difference, while IAA/HD/II-F do not; and MME/IFD are pairwise measures. 

\noindent \textbf{Range of scores.} We identify three issues on the score range of the \textsc{Fair+Rel} measures: (1) extremely small scales for several joint measures; (2) scale mismatch between single-aspect measures and joint measures; and (3) scale mismatch between joint measures. 
About (1), for all datasets and models, several \down \textsc{Fair+Rel} scores are extremely small ($\leq10^{-3}$), and these scores do not allow to distinguish across models per dataset. 
For example, IFD$_{\times}$ is always close to 0 across all datasets, as the term Eq.~\eqref{eq:ifd_j_x} is often 0 due to the low number of relevant items per user.\footnote{For all four datasets, the median number of relevant items per user is at most 46.} For MME and II-F/AI-F, Eq.~\eqref{eq:imp} and Eq.~\eqref{eq:eui-star} often result in 0 for the same reason as IFD$_{\times}$. 
About (2), while the above \textsc{Fair+Rel} scores differ in the fourth or later decimal point, the differences in the \textsc{Rel} and \textsc{Fair} scores are in the second decimal point or before. E.g., the NDCG (\textsc{Rel} score) of MultiVAE-CM and NCL for Lastfm differs by $\sim$0.16 and their Jain (\textsc{Fair} score) differs by $\sim$0.14. These examples imply non-negligible differences, but the joint scores of IAA/IFD$_{\times}$/MME/II-F/AI-F only differ by $\leq 10^{-3}$, which may seem negligible.\footnote{We use NDCG and Jain as they are more sensitive to changes than HR and QF.} 
These inconsistencies in the difference of magnitude make the scores hard to understand. 
About (3), we see large gaps in the score range of all joint measures, e.g., between IWO, HD, and AI-F, despite all of them being lower-is-better measures. E.g., in ML-10M, \down IWO $\approx 1$ (very unfair) based on its theoretical $[0,1]$-range, \down HD is about a quarter of the \down IWO score (somewhat fair), while \down AI-F $\approx 0$ (extremely fair). This discrepancy causes confusion in score interpretation. 

Finally, we group all \textsc{Fair+Rel} measures into 3 clusters: 
(i) IAA /HD/II-F, which align more with \textsc{Rel} measures; 
(ii) IFD/MME/AI-F, which align more with \textsc{Fair} measures; and
(iii) IBO/IWO, which do not consistently align with any single-aspect measure. Within the same cluster, especially in (i), measures often have large differences in their score ranges (up to $\Delta\approx0.7$).

\subsection{Correlation between measures (RQ1 \& RQ2)}
\label{ss:corr}

\begin{figure*}
    \centering
    \includegraphics[width=0.9\textwidth, trim=0.25cm 0.6cm 0.3cm 0.3cm clip=True]{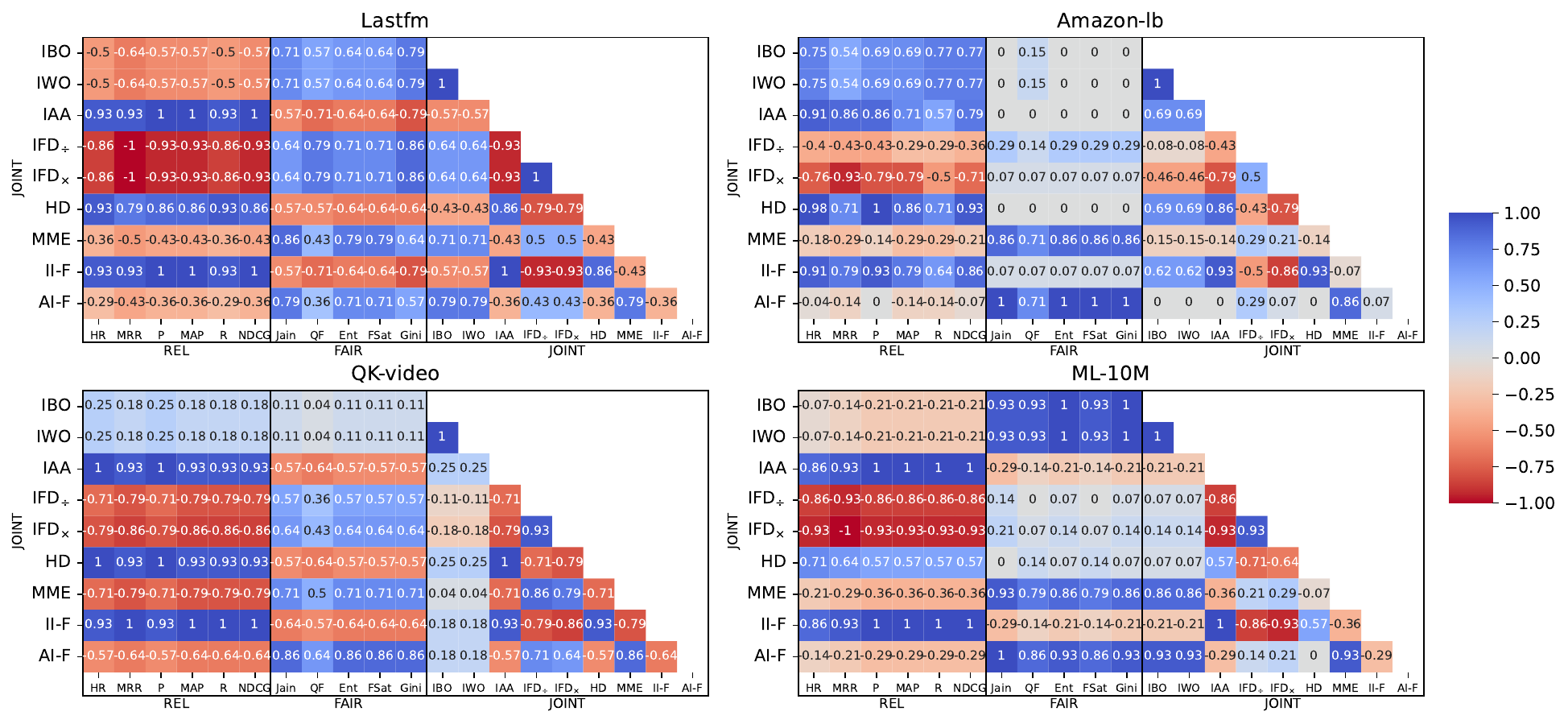}
    \caption{Kendall's $\tau$ correlation between joint \textsc{Fair+Rel} measures, \textsc{Rel}, and \textsc{Fair} measures.}
    \label{fig:corr-grid}
\end{figure*}

We compute Kendall's $\tau$ correlation between the orderings of the recommenders produced by the scores of each measure, to study how much the \textsc{Fair+Rel} measures agree among themselves but also with \textsc{Rel}-only and \textsc{Fair}-only measures, when ranking recommenders 
(Fig.~\ref{fig:corr-grid}). As \textsc{Rel} and \textsc{Fair} measures do not always correlate strongly with each other \cite{Rampisela2023EvaluationStudy}, we do not expect \textsc{Fair+Rel} measures to correlate strongly with either \textsc{Rel} or \textsc{Fair} measures.

\noindent\textbf{RQ1. Agreement of joint and single-aspect measures.} Overall, there is no consistent correlation between \textsc{Rel} and \textsc{Fair+Rel} measures. IBO/IWO's correlations vary wildly ( $\tau \in [-0.64, 0.77]$); 
IAA, HD, and II-F have moderate-to-strong positive correlations ($\tau \in [0.57, 1]$); 
IFD and MME have weak-to-strong negative correlations ($\tau \in [-1, -0.29]$ for IFD and $\tau \in [-0.79, -0.14]$ for MME); and AI-F has non-positive correlations ($\tau \in [-0.64,0]$). 

The correlations between \textsc{Fair} and \textsc{Fair+Rel} measures are inconsistent. The correlations of IBO/IWO vary largely again, albeit less than with \textsc{Rel} measures. IAA/HD/II-F have two distinct trends across groups of datasets: they have negative moderate-to-strong correlations ($\tau \in [-0.79,-0.57]$) for Lastfm and QK-video, but weak correlations for Amazon-lb and ML-10M ($\tau \in [-0.29,0.14]$). Similarly, IFD has high correlations for Lastfm and QK-video (except with QF for QK-video), $\tau \in [0.57, 0.86]$, and weak or zero correlations for the other datasets ($\tau\in [0, 0.29]$). Conversely, MME and AI-F have strong correlations except with QF for Lastfm ($\tau \in [0.5,1]$). 

Note that \textsc{Fair+Rel} measures strongly agreeing with \textsc{Rel} measures do not always strongly disagree with \textsc{Fair} measures, and vice versa. E.g., IAA/HD/II-F strongly correlates with \textsc{Rel} measures for Amazon-lb, but they correlate weakly with \textsc{Fair} measures. 

\noindent\textbf{RQ2. Agreement between joint measures.} Overall we find that the three clusters of joint measures identified in $\S$\ref{ss:performance} show strong positive correlations between measures inside the same cluster and strong negative correlations between measures from different clusters. E.g., IBO always perfectly correlates with IWO, due to their similar formulation. 
IAA, HD, and II-F agree strongly with one another, $\tau \in [0.57,1]$. IFD$_{\div}$ correlates highly with IFD$_{\times}$, $\tau \in [0.5,1]$, as their formulations are similar. MME always agrees strongly with AI-F, $\tau \in [0.79,0.93]$. IFD sometimes has moderate-to-strong correlations with MME and AI-F, $\tau \in [0.43,0.86]$ for Lastfm and QK-video, but the correlations are weaker for Amazon-lb and ML-10M, $\tau \in [0.07,0.29]$. In contrast, IAA/HD/II-F strongly disagrees with IFD, $\tau \in [-0.71,-0.5]$ except for IFD$_{\div}$ in Amazon-lb ($\tau=-0.43$).

Based on the above, we conclude that: IBO/IWO has inconsistent relationships with single-aspect and joint measures; IAA/HD/II-F do not align with fairness; and IFD/MME/AI-F highly disagree with relevance (even if IFD sometimes disagrees with \textsc{Fair} measures too). Among the joint measures, IBO/IWO weakly correlate with the single-aspect measures for QK-video, and similarly with IFD$_{\div}$ for Amazon-lb, but this is not consistent. We thus argue that no joint measures reliably account for both relevance and fairness.

\subsection{
Measure sensitivity at different ranks (RQ3)
}
\label{ss:sliding}

\begin{figure*}
    \centering
    \includegraphics[width=0.9\textwidth, trim=0cm 0cm 0cm 0.2cm, clip=True]{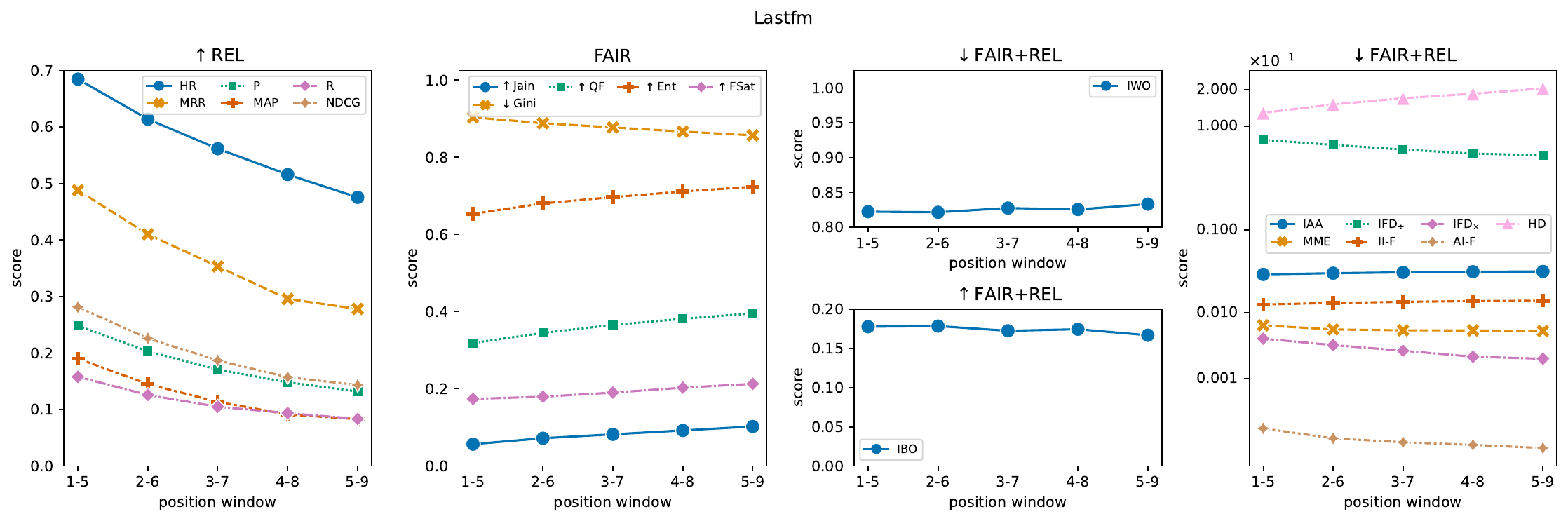}
    \includegraphics[width=0.9\textwidth]{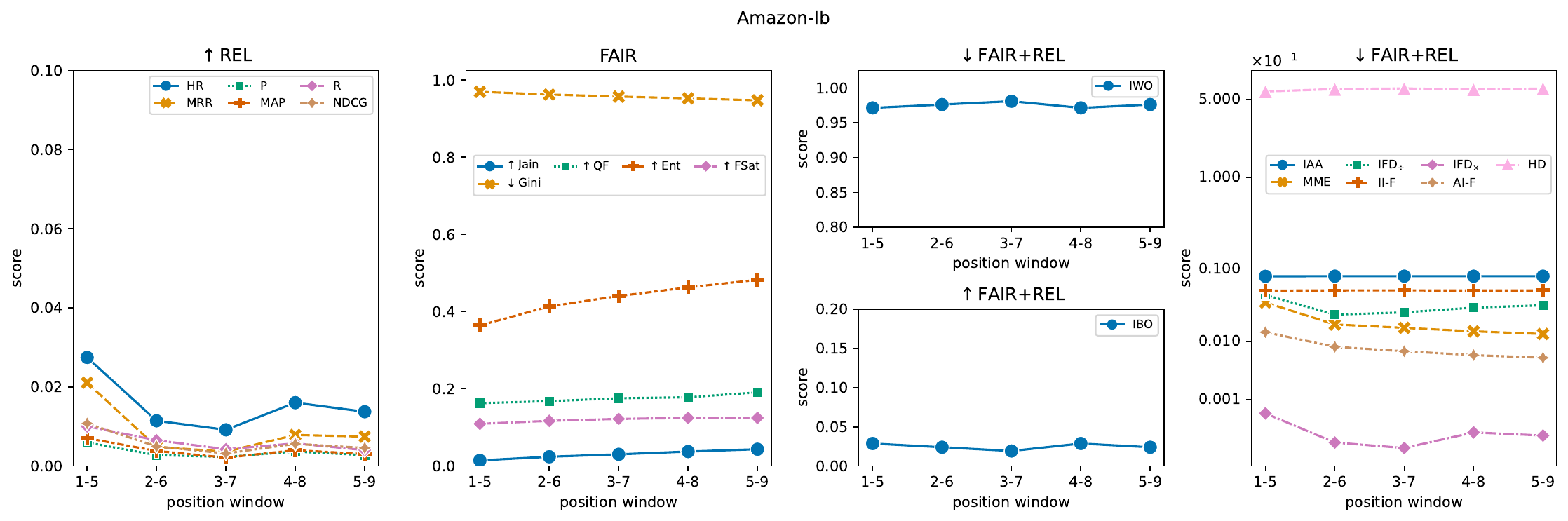}
    \includegraphics[width=0.9\textwidth, trim=0cm 0.5cm 0cm 0cm]{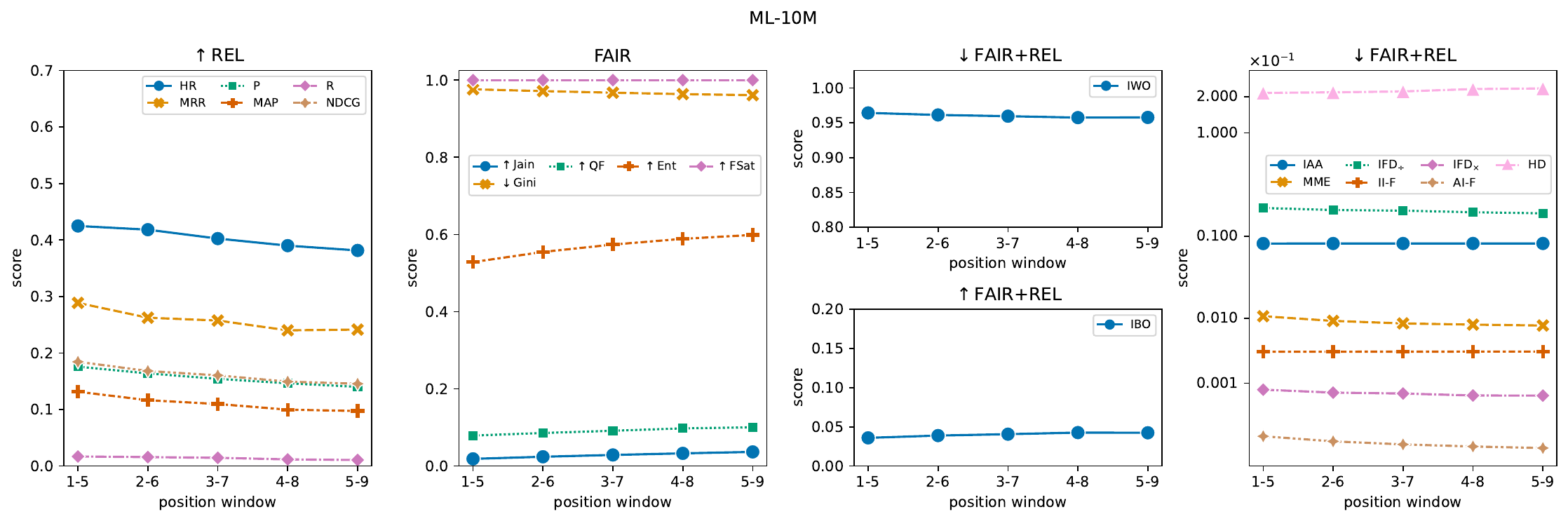}
    \caption{Sliding window evaluation (\(\mathbf{k=5}\)) of NCL for Lastfm, Amazon-lb, and ML-10M. The last column is in exponential scale.}
    \label{fig:sliding}
\end{figure*}

We now study how sensitive the joint measures are at decreasing rank positions, compared to \textsc{Rel} and \textsc{Fair} measures. When moving down the rank, \textsc{Rel} scores are known to decrease while \textsc{Fair} scores are known to improve \cite{Rampisela2023EvaluationStudy}. For this analysis, we use only the runs of the non-reranked NCL model as it generally has the best \textsc{Rel} scores. We compute all measures at $k=5$ for each sliding window, where the windows consist of items at decreasing rank positions: 1--5, 2--6, $\dots$, 5--9. Fig.~\ref{fig:sliding} shows the results for Lastfm, Amazon-lb, and ML-10M, which represent the overall trends in all our datasets; results for QK-video are shown in the appendix (in our code repository). 

We find that, as expected, as we move down the rank, \textsc{Rel} overall decreases and \textsc{Fair} improves. However, the joint measures are notably less sensitive to changes in rank position. 
Changes with decreasing rank position in the single-aspect scores are up two magnitudes greater than in the joint measures, and the latter do not reflect these differences to the same scale. We posit that the insensitivity is due to the effect of changing relevance being masked by that of fairness and vice versa. This masking makes the scores hard to interpret. Further, the very small scores of \down IAA, IFD$_{\times}$, MME, II-F, and AI-F 
imply extremely fair recommendations (we explain the reasons for this in $\S$\ref{ss:performance}), even if \textsc{Rel} and \textsc{Fair} scores are low. Thus, these joint measures do not account well for relevance and fairness simultaneously. Last, we note that as we move down the rank, IAA/HD/II-F worsen, IFD/MME/AI-F improve, and IBO/IWO are inconsistent across datasets. This follows the three groups of joint measures discussed above. 

\subsection{Artificial insertion of items (RQ4)}
\label{ss:insert}

\begin{figure}
    \centering
    \includegraphics[width=0.8\columnwidth, trim=1.5cm 0.5cm 1.5cm 0cm]{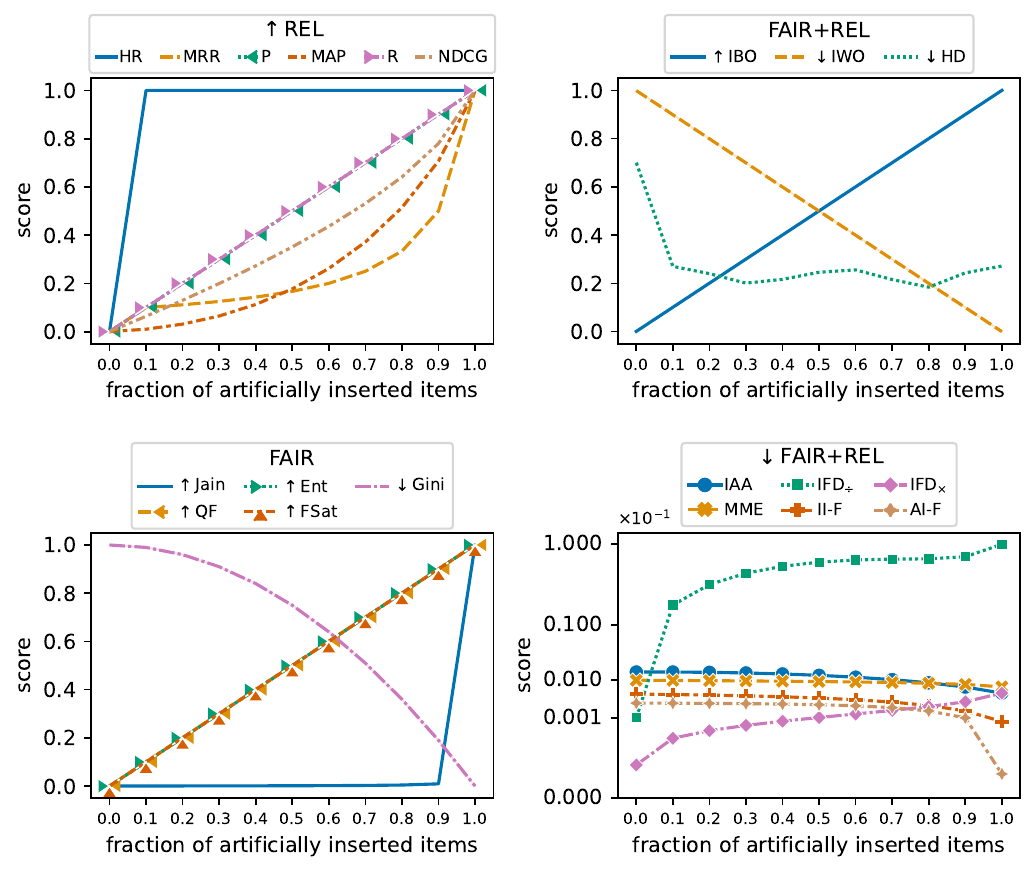}
    \caption{Artificial insertion of items with \(\mathbf{m=1000}\) (users).}
    \label{fig:insert}
\end{figure}

Lastly, we study how sensitive the joint measures are to different proportions of relevant items and item fairness in the ranking. Assessing this sensitivity is important as it affects score interpretation; if a joint measure is unresponsive to significant changes in relevance and fairness distribution, its score may not reflect both the fairness and relevance of the recommendations accurately. 

We start with a recommendation list having the worst \textsc{Rel} and \textsc{Fair} scores, and gradually insert more relevant and fair items to it (we explain `fair items' below). 
We observe how the joint measures respond to these changes, compared to the \textsc{Rel} and \textsc{Fair} measures.

We cannot use real-life datasets for this analysis, so we build a synthetic dataset with $m=1000$ and $n=10000$, and artificially generate rankings of items per user, as per \cite{Rampisela2023EvaluationStudy}. 
The initial ranking contains the same $k=10$ items for all users, to whom these items are irrelevant, except for one user.\footnote{This is to keep the number of items exactly $km$.} 
In each iteration, an item from the bottom of each user's top $k$ ranking is replaced by a relevant item having less exposure (hence more fair). The final ranking thus contains $km$ unique items across all users, where each item is relevant only to the user that receives that item in the top $k$. We expect all measures to initially score the worst possible, and then gradually improve as more relevant and fair items enter the ranks. 

Fig.~\ref{fig:insert} shows the results of this analysis. Overall, we see that most joint measures are not very sensitive to changes in \textsc{Rel} and \textsc{Fair} scores, i.e., they may vary, but negligibly. This verifies the scale mismatch between most joint measures and the single-aspect measures observed in $\S$\ref{ss:performance} \& $\S$\ref{ss:sliding}. 
While the overall change is negligible for most measures, a common observation between the joint measures is that their scores become (slightly) better as \textsc{Rel} and \textsc{Fair} scores improve. An exception to this is IFD. This is because IFD measures fairness based on the pairwise difference in the combined value of exposure and relevance. Thus, when the relevant items start to be moved to the top $k$, the gap between the exposure weight of relevant items in and outside the top $k$ increases, and so does unfairness. Among joint measures that (slightly) improve with more insertion, there are also differences. 
IBO/IWO improve linearly; as both measures are percentages of items, the change is proportional to the amount of inserted items. HD also improves, but its improvement fluctuates due to randomness introduced by the unstable sort in the computation, as per the original implementation in \cite{Jeunen2021Top-KExposure}. \down IAA/IFD$_{\times}$/MME/II-F/AI-F improve non-linearly. However, their scores are extremely close to 0, i.e., on the scale of $10^{-3}$ or less. The lower bound of the measure is 0, hence these small scores indicate that the recommendation is close to the fairest, even at the start of the process where the \textsc{Rel} and \textsc{Fair} scores are the worst in the entire progression. These joint measures are also rather insensitive to changes in \textsc{Rel} and \textsc{Fair} scores. Here, their score range is (0, 0.0015), while the range of \textsc{Rel}, \textsc{Fair}, and IBO/IWO scores is [0,1]. 

\section{Related work}\label{s:prevwork}

\noindent\textbf{Fairness evaluation in RSs}. Among prior work on fairness evaluation in RSs \cite{Wang2022,Amigo2023ASystems,Zehlike2022FairnessSystems,Raj2022MeasuringResults, Rampisela2023EvaluationStudy}, our study is close to
\citeauthor{Amigo2023ASystems}~\cite{Amigo2023ASystems}, who study RS relevance and fairness for groups/individuals and between
items and users. Yet, the focus of our work, i.e., individual item fairness, is not covered in \cite{Amigo2023ASystems}. \citeauthor{Raj2022MeasuringResults}~\cite{Raj2022MeasuringResults} overview evaluation measures for item group fairness. Their study includes the IAA measure as a group fairness measure (whereas we focus on individual item fairness). Lastly, 
\citeauthor{Rampisela2023EvaluationStudy}~\cite{Rampisela2023EvaluationStudy} survey individual item fairness measures that are exclusively linked with fairness and identify the limitations within them, while we focus on  measures that jointly account for both fairness and relevance. 

\noindent\textbf{Joint measures of relevance and fairness}. Outside the strict domain of individual item fairness for RS, there exist other measures that quantify relevance and fairness jointly: \citeauthor{Gao2022FAIR:Evaluation}~\cite{Gao2022FAIR:Evaluation} present a measure combining KL-divergence and IDCG to jointly quantify relevance and group fairness in IR \cite{Gao2022FAIR:Evaluation}. In \cite{Xu2023P-MMF:System}, utility and provider fairness in RSs are simultaneously evaluated with a weighted sum between relevance and fairness. Another approach used in \cite{Garcia-Soriano2021Maxmin-FairConstraints} to evaluate individual fairness in ranking is to compare item position based on ground truth relevance against its position in system-produced rankings. None of the joint measures in our work is a combination of two single-aspect measures as in \cite{Gao2022FAIR:Evaluation} or in the form of weighted sum as in \cite{Xu2023P-MMF:System}. The measure in \cite{Garcia-Soriano2021Maxmin-FairConstraints} is similar to HD \cite{Jeunen2021Top-KExposure}. However, we do not use it in our work because it was not defined for RS fairness, and considerable modifications and assumptions are required prior to using it to evaluate RS fairness.

\section{Appropriate usage of joint measures}
\label{s:discussion}

We find that joint measures of relevance and fairness (1) tend to align differently with single-aspect measures; 
(2) most of them consistently score almost perfect fairness, even when recommendations are highly irrelevant and unfair based on single-aspect measures; and (3) are rather unresponsive to changes in the recommendation relevance and fairness, especially compared to single-aspect measures. 
Next, we suggest how to best use these joint measures.

\noindent
\textbf{Avoid using similar joint measures}.
In $\S$\ref{ss:corr} \& $\S$\ref{ss:sliding} we find three groups of similar joint measures: (i) IAA/HD/II-F, (ii) IFD/MME/AI-F, and (iii) IBO/IWO. Only one measure per group should be used. Yet, considering that typically recommendations are evaluated with \textsc{Rel} measures, we discourage using measures in (i), as they are highly aligned with \textsc{Rel} measures. Measures in (ii) correlate strongly with \textsc{Fair} measures, and can be viable options, and likewise for measures in (iii) that do not consistently correlate with single-aspect measures. However, we argue that measures in (iii) are more useful than those in (ii). Measures in (ii) can be replaced by \textsc{Fair} measures, which are faster to compute and do not need complete relevance judgements, while still achieving highly similar conclusions.

\noindent
\textbf{Be aware of the unintuitive or inconsistent behaviour, insensitivity, and computational complexity of the measures}. 
We recommend that practitioners be aware of the measure limitations in groups (ii) and (iii). Specifically, both IFD versions worsen with higher percentages of jointly relevant and fair recommendations, while the opposite should happen in a healthy measure. 
IFD$_{\div}$ is unaffected by different cut-off $k$ values, as its original formulation only considers full rankings, so it should not be used when different $k$ matters. MME is costly to compute as it is a pairwise measure ($\sim$30 mins for the larger datasets), and the same applies to \down IFD$_{\times}$, albeit to a lesser extent. Further, IFD$_{\times}$/MME/AI-F tend to have extremely small scores, which are therefore hard to interpret and discriminate across runs. They are also rather insensitive to changes reflected by single-aspect measures, meaning that their overall expressiveness is limited. IBO/IWO is sensitive in this aspect. Considering the limitations and the redundancy between measures, IBO/IWO seem to be the most viable measure out of the existing ones, but it is the least consistent between all other measures, due to varying alignments for different datasets, so it should be interpreted cautiously. 

\noindent
\textbf{Avoid score misinterpretation in measures with small empirical scales.}
Due to the small empirical scales of \down IAA/IFD/MME/II-F/AI-F, 
their scores tend not to represent fairness, or relevance and fairness jointly, i.e., scoring very close to 0 even if the recommendation is very irrelevant and unfair based on \textsc{Rel} and \textsc{Fair} measures. Moreover, different systems can have very similar scores which do not translate to similar performance. 
E.g., two models differing in scores by only 0.001 can be interpreted as performing the same, even though the measure has a small empirical range and is not very sensitive to begin with. This issue can be fixed via apriori/posthoc normalisation based on experimental values of the measures \cite{Wu2022JointRecommendation}.  

\noindent\textbf{Measure fairness separately from relevance}. 
As most joint measures (IAA/IFD/MME/II-F/AI-F) are difficult to interpret because their scores tend to be compressed in a very low range, and are also rather insensitive to changes in fairness and relevance, we recommend measuring individual item fairness and relevance separately. Otherwise, the joint scores can be close to the theoretical fairest value even if \textsc{Rel} and \textsc{Fair} scores are low ($\S$\ref{ss:insert}). 
Overall, the above joint measures have unreliable scores, are not as sensitive as the \textsc{Fair} measures, and are subject to more under/overestimation of fairness than \textsc{Fair} measures which have more consistent empirical range. The remaining joint measures are not reliable either: IBO/IWO aligns inconsistently to the single-aspect measures, while HD is almost always consistent with \textsc{Rel} measures and thus does not add another dimension of fairness measurement. It is also unstable due to sorting of items with identical relevance level. 

Overall, the joint measures cannot be compared easily as they have different scales, and they quantify two aspects that are hard to combine due to mismatching scales. The measures tend to correlate highly with either \textsc{Rel} or \textsc{Fair} measures, instead of having a good balance between them. As such, optimising for a joint measure directly may not result in a simultaneously optimal recommendation based on \textsc{Rel} and \textsc{Fair} scores. Another obstacle in measuring fairness is the need to consider user-item relevance in the entire dataset (not just the recommended items), which can be an issue with extremely sparse datasets. It is thus inherently difficult to devise a measure that can jointly quantify relevance and fairness.

\section{Conclusions and future work}

We presented a novel empirical study on the properties of all evaluation measures that jointly account for individual item fairness and relevance in recommender systems. We found that out of 9 joint measures, 3 align with traditional relevance-only measures, 4 agree more with fairness-only measures, and the rest behave inconsistently. We also found that only a few joint measures are sensitive to a simultaneous decrease in relevance and increased fairness in the recommendation. Most surprisingly, nearly all joint measures are almost unresponsive to increases in relevance and fairness. Even worse, the majority tend to compress scores at the low end of their range, giving the illusion of an extremely fair recommendation, even when the relevance- and fairness-only scores are close to the theoretical worst value. Based on these findings, we formulated recommendations on the appropriate usage of these measures. 

Future work includes improving the design of joint measures by addressing or mitigating the limitations of the current measures outlined, to have a single score that reflects recommendation relevance and fairness more accurately and in a more balanced way. The individual fairness and relevance measures can also be optimised jointly with multi-objective approach, to obtain both fair and relevant recommendations. Future work can also investigate whether the findings hold when the models are directly optimised for fairness, or when different family of models are used.


\begin{acks}
Funded by Algorithms, Data \& Democracy (Villum \& Velux funds). 
\end{acks}

\printbibliography

\end{document}

%% file: tab/base_table.tex
\begin{table*}
\caption{Relevance (\textsc{Rel}), fairness (\textsc{Fair}), and joint \textsc{Fair+Rel} scores at \(\mathbf{k=10}\) without and with re-ranking the top \(\mathbf{k'=25}\) items using COMBMNZ (CM). Bold marks the most relevant/fair score per measure. The score 0.000 does not mean the scores are exactly 0; this is due to the measures having small scores (\(\mathbf{<10^{-3}}\)) and rounding to 3 d.p.
}
\label{tab:base-rerank-all}
\begin{minipage}[t]{.49\textwidth}
\resizebox{0.99\linewidth}{!}{
\begin{tabular}[t]{lll*{2}{r}|*{2}{r}|*{2}{r}|*{2}{r}}
\toprule
 &  & model & \multicolumn{2}{c|}{ItemKNN} & \multicolumn{2}{c|}{BPR} & \multicolumn{2}{c|}{MultiVAE} & \multicolumn{2}{c}{NCL} \\ 
\midrule
 &  & re-ranker & - & CM & - & CM & - & CM & - & CM \\
\midrule
\multirow[c]{21}{*}{\rotatebox[origin=c]{90}{Lastfm}} & \multirow[c]{6}{*}{\rotatebox[origin=c]{90}{\textsc{Rel}}} & $\uparrow$ $\text{HR}$ & 0.765 & 0.581 & 0.773 & 0.587 & 0.778 & 0.523 & \bfseries 0.793 & 0.571 \\
 &  & $\uparrow$ $\text{MRR}$ & 0.484 & 0.270 & 0.492 & 0.280 & 0.476 & 0.232 & \bfseries 0.503 & 0.260 \\
 &  & $\uparrow$ $\text{P}$ & 0.172 & 0.089 & 0.178 & 0.092 & 0.176 & 0.076 & \bfseries 0.184 & 0.087 \\
 &  & $\uparrow$ $\text{MAP}$ & 0.137 & 0.053 & 0.141 & 0.058 & 0.138 & 0.045 & \bfseries 0.148 & 0.050 \\
 &  & $\uparrow$ $\text{R}$ & 0.218 & 0.114 & 0.224 & 0.119 & 0.224 & 0.098 & \bfseries 0.234 & 0.110 \\
 &  & $\uparrow$ $\text{NDCG}$ & 0.245 & 0.119 & 0.252 & 0.126 & 0.247 & 0.102 & \bfseries 0.261 & 0.115 \\
\cline{2-11}
 & \multirow[c]{5}{*}{\rotatebox[origin=c]{90}{\textsc{Fair}}} & $\uparrow$ $\text{Jain}$ & 0.042 & 0.094 & 0.058 & 0.140 & 0.097 & \bfseries 0.222 & 0.082 & 0.215 \\
 &  & $\uparrow$ $\text{QF}$ & 0.474 & \bfseries 0.679 & 0.362 & 0.528 & 0.517 & 0.678 & 0.453 & 0.657 \\
 &  & $\uparrow$ $\text{Ent}$ & 0.589 & 0.735 & 0.610 & 0.740 & 0.707 & \bfseries 0.826 & 0.671 & 0.810 \\
 &  & $\uparrow$ $\text{FSat}$ & 0.129 & 0.216 & 0.147 & 0.228 & 0.202 & \bfseries 0.321 & 0.178 & 0.286 \\
 &  & $\downarrow$ $\text{Gini}$ & 0.904 & 0.790 & 0.910 & 0.818 & 0.839 & \bfseries 0.696 & 0.872 & 0.728 \\
\cline{2-11}
 & \multirow[c]{9}{*}{\rotatebox[origin=c]{90}{\textsc{Fair+Rel}}} & $\uparrow$ $\text{IBO}$ & 0.209 & 0.256 & 0.208 & 0.253 & 0.261 & 0.278 & 0.242 & \bfseries 0.292 \\
 &  & $\downarrow$ $\text{IWO}$ & 0.791 & 0.744 & 0.792 & 0.747 & 0.739 & 0.722 & 0.758 & \bfseries 0.708 \\
 &  & $\downarrow$ $\text{IAA}$ & 0.004 & 0.004 & 0.004 & 0.004 & 0.004 & 0.004 & \bfseries 0.004 & 0.004 \\
 &  & $\downarrow$ $\text{IFD}_{\div}$ & 0.074 & 0.053 & 0.075 & 0.054 & 0.073 & \bfseries 0.049 & 0.076 & 0.052 \\
 &  & $\downarrow$ $\text{IFD}_{\times}$ & 0.000 & 0.000 & 0.000 & 0.000 & 0.000 & \bfseries 0.000 & 0.000 & 0.000 \\
 &  & $\downarrow$ $\text{HD}$ & 0.099 & 0.177 & 0.104 & 0.174 & 0.095 & 0.203 & \bfseries 0.092 & 0.177 \\
 &  & $\downarrow$ $\text{MME}$ & 0.001 & 0.001 & 0.001 & 0.001 & 0.001 & 0.001 & 0.001 & \bfseries 0.001 \\
 &  & $\downarrow$ $\text{II-F}$ & 0.001 & 0.002 & 0.001 & 0.002 & 0.001 & 0.002 & \bfseries 0.001 & 0.002 \\
 &  & $\downarrow$ $\text{AI-F}$ & 0.000 & 0.000 & 0.000 & 0.000 & 0.000 & 0.000 & 0.000 & \bfseries 0.000 \\
\cline{1-11}
\multirow[c]{21}{*}{\rotatebox[origin=c]{90}{Amazon-lb}} & \multirow[c]{6}{*}{\rotatebox[origin=c]{90}{\textsc{Rel}}} & $\uparrow$ $\text{HR}$ & \bfseries 0.046 & 0.016 & 0.011 & 0.021 & 0.039 & 0.014 & 0.034 & 0.011 \\
 &  & $\uparrow$ $\text{MRR}$ & 0.020 & 0.011 & 0.003 & 0.007 & \bfseries 0.023 & 0.004 & 0.022 & 0.003 \\
 &  & $\uparrow$ $\text{P}$ & \bfseries 0.005 & 0.002 & 0.001 & 0.002 & 0.004 & 0.002 & 0.004 & 0.001 \\
 &  & $\uparrow$ $\text{MAP}$ & 0.006 & 0.004 & 0.002 & 0.004 & 0.006 & 0.003 & \bfseries 0.006 & 0.001 \\
 &  & $\uparrow$ $\text{R}$ & \bfseries 0.013 & 0.005 & 0.005 & 0.010 & 0.010 & 0.008 & 0.012 & 0.003 \\
 &  & $\uparrow$ $\text{NDCG}$ & \bfseries 0.011 & 0.005 & 0.003 & 0.006 & 0.010 & 0.004 & 0.011 & 0.002 \\
\cline{2-11}
 & \multirow[c]{5}{*}{\rotatebox[origin=c]{90}{\textsc{Fair}}} & $\uparrow$ $\text{Jain}$ & 0.271 & \bfseries 0.431 & 0.223 & 0.359 & 0.035 & 0.097 & 0.026 & 0.080 \\
 &  & $\uparrow$ $\text{QF}$ & \bfseries 0.650 & 0.612 & 0.549 & 0.594 & 0.222 & 0.286 & 0.229 & 0.310 \\
 &  & $\uparrow$ $\text{Ent}$ & 0.802 & \bfseries 0.839 & 0.747 & 0.809 & 0.418 & 0.558 & 0.371 & 0.534 \\
 &  & $\uparrow$ $\text{FSat}$ & 0.370 & \bfseries 0.438 & 0.314 & 0.376 & 0.114 & 0.152 & 0.091 & 0.138 \\
 &  & $\downarrow$ $\text{Gini}$ & 0.665 & \bfseries 0.598 & 0.747 & 0.660 & 0.949 & 0.899 & 0.959 & 0.910 \\
\cline{2-11}
 & \multirow[c]{9}{*}{\rotatebox[origin=c]{90}{\textsc{Fair+Rel}}} & $\uparrow$ $\text{IBO}$ & \bfseries 0.062 & 0.029 & 0.019 & 0.038 & 0.029 & 0.029 & 0.038 & 0.024 \\
 &  & $\downarrow$ $\text{IWO}$ & \bfseries 0.938 & 0.971 & 0.981 & 0.962 & 0.971 & 0.971 & 0.962 & 0.976 \\
 &  & $\downarrow$ $\text{IAA}$ & \bfseries 0.011 & 0.011 & 0.011 & 0.011 & 0.011 & 0.011 & 0.011 & 0.011 \\
 &  & $\downarrow$ $\text{IFD}_{\div}$ & 0.005 & 0.003 & 0.003 & \bfseries 0.002 & 0.005 & 0.002 & 0.005 & 0.003 \\
 &  & $\downarrow$ $\text{IFD}_{\times}$ & 0.000 & 0.000 & 0.000 & 0.000 & 0.000 & 0.000 & 0.000 & \bfseries 0.000 \\
 &  & $\downarrow$ $\text{HD}$ & \bfseries 0.580 & 0.630 & 0.661 & 0.626 & 0.597 & 0.653 & 0.598 & 0.667 \\
 &  & $\downarrow$ $\text{MME}$ & 0.001 & \bfseries 0.001 & 0.001 & 0.001 & 0.003 & 0.001 & 0.004 & 0.001 \\
 &  & $\downarrow$ $\text{II-F}$ & \bfseries 0.006 & 0.006 & 0.006 & 0.006 & 0.006 & 0.006 & 0.006 & 0.006 \\
 &  & $\downarrow$ $\text{AI-F}$ & 0.000 & \bfseries 0.000 & 0.000 & 0.000 & 0.001 & 0.000 & 0.002 & 0.000 \\
\bottomrule
\end{tabular}
}
\end{minipage}%
\noindent\begin{minipage}[t]{.49\textwidth}
\resizebox{0.995\linewidth}{!}{
\begin{tabular}[t]{lll*{2}{r}|*{2}{r}|*{2}{r}|*{2}{r}}
\toprule
 &  & model & \multicolumn{2}{c|}{ItemKNN} & \multicolumn{2}{c|}{BPR} & \multicolumn{2}{c|}{MultiVAE} & \multicolumn{2}{c}{NCL} \\ 
\midrule
 &  & re-ranker & - & CM & - & CM & - & CM & - & CM \\
\midrule
\multirow[c]{21}{*}{\rotatebox[origin=c]{90}{QK-video}} & \multirow[c]{6}{*}{\rotatebox[origin=c]{90}{\textsc{Rel}}} & $\uparrow$ $\text{HR}$ & 0.040 & 0.047 & 0.099 & 0.045 & 0.109 & 0.061 & \bfseries 0.130 & 0.077 \\
 &  & $\uparrow$ $\text{MRR}$ & 0.013 & 0.013 & 0.039 & 0.015 & 0.039 & 0.021 & \bfseries 0.048 & 0.024 \\
 &  & $\uparrow$ $\text{P}$ & 0.004 & 0.005 & 0.011 & 0.005 & 0.012 & 0.006 & \bfseries 0.014 & 0.008 \\
 &  & $\uparrow$ $\text{MAP}$ & 0.005 & 0.005 & 0.017 & 0.006 & 0.018 & 0.009 & \bfseries 0.022 & 0.010 \\
 &  & $\uparrow$ $\text{R}$ & 0.014 & 0.019 & 0.043 & 0.019 & 0.051 & 0.027 & \bfseries 0.061 & 0.033 \\
 &  & $\uparrow$ $\text{NDCG}$ & 0.009 & 0.010 & 0.029 & 0.011 & 0.031 & 0.016 & \bfseries 0.038 & 0.019 \\
\cline{2-11}
 & \multirow[c]{5}{*}{\rotatebox[origin=c]{90}{\textsc{Fair}}} & $\uparrow$ $\text{Jain}$ & 0.483 & \bfseries 0.589 & 0.081 & 0.379 & 0.012 & 0.032 & 0.020 & 0.071 \\
 &  & $\uparrow$ $\text{QF}$ & \bfseries 0.901 & 0.790 & 0.625 & 0.823 & 0.100 & 0.163 & 0.201 & 0.365 \\
 &  & $\uparrow$ $\text{Ent}$ & 0.933 & \bfseries 0.937 & 0.755 & 0.903 & 0.420 & 0.547 & 0.507 & 0.674 \\
 &  & $\uparrow$ $\text{FSat}$ & 0.443 & \bfseries 0.547 & 0.212 & 0.382 & 0.052 & 0.090 & 0.077 & 0.150 \\
 &  & $\downarrow$ $\text{Gini}$ & 0.472 & \bfseries 0.442 & 0.807 & 0.570 & 0.982 & 0.959 & 0.966 & 0.902 \\
\cline{2-11}
 & \multirow[c]{9}{*}{\rotatebox[origin=c]{90}{\textsc{Fair+Rel}}} & $\uparrow$ $\text{IBO}$ & 0.033 & 0.038 & \bfseries 0.054 & 0.036 & 0.031 & 0.036 & 0.043 & \bfseries 0.054 \\
 &  & $\downarrow$ $\text{IWO}$ & 0.967 & 0.962 & \bfseries 0.946 & 0.964 & 0.969 & 0.964 & 0.957 & \bfseries 0.946 \\
 &  & $\downarrow$ $\text{IAA}$ & 0.001 & 0.001 & 0.001 & 0.001 & 0.001 & 0.001 & \bfseries 0.001 & 0.001 \\
 &  & $\downarrow$ $\text{IFD}_{\div}$ & 0.009 & \bfseries 0.007 & 0.014 & 0.008 & 0.014 & 0.009 & 0.015 & 0.010 \\
 &  & $\downarrow$ $\text{IFD}_{\times}$ & 0.000 & \bfseries 0.000 & 0.000 & 0.000 & 0.000 & 0.000 & 0.000 & 0.000 \\
 &  & $\downarrow$ $\text{HD}$ & 0.576 & 0.560 & 0.490 & 0.565 & 0.478 & 0.535 & \bfseries 0.457 & 0.519 \\
 &  & $\downarrow$ $\text{MME}$ & 0.000 & \bfseries 0.000 & 0.000 & 0.000 & 0.000 & 0.000 & 0.000 & 0.000 \\
 &  & $\downarrow$ $\text{II-F}$ & 0.001 & 0.001 & 0.001 & 0.001 & 0.001 & 0.001 & \bfseries 0.001 & 0.001 \\
 &  & $\downarrow$ $\text{AI-F}$ & 0.000 & \bfseries 0.000 & 0.000 & 0.000 & 0.000 & 0.000 & 0.000 & 0.000 \\
\cline{1-11}
\multirow[c]{21}{*}{\rotatebox[origin=c]{90}{ML-10M}} & \multirow[c]{6}{*}{\rotatebox[origin=c]{90}{\textsc{Rel}}} & $\uparrow$ $\text{HR}$ & 0.487 & 0.443 & 0.512 & 0.386 & 0.417 & 0.387 & \bfseries 0.521 & 0.402 \\
 &  & $\uparrow$ $\text{MRR}$ & 0.282 & 0.225 & 0.299 & 0.185 & 0.237 & 0.191 & \bfseries 0.302 & 0.203 \\
 &  & $\uparrow$ $\text{P}$ & 0.137 & 0.105 & 0.146 & 0.088 & 0.107 & 0.096 & \bfseries 0.154 & 0.094 \\
 &  & $\uparrow$ $\text{MAP}$ & 0.089 & 0.060 & 0.095 & 0.047 & 0.067 & 0.054 & \bfseries 0.101 & 0.052 \\
 &  & $\uparrow$ $\text{R}$ & 0.022 & 0.018 & 0.025 & 0.012 & 0.020 & 0.016 & \bfseries 0.026 & 0.013 \\
 &  & $\uparrow$ $\text{NDCG}$ & 0.150 & 0.113 & 0.160 & 0.092 & 0.119 & 0.100 & \bfseries 0.167 & 0.100 \\
\cline{2-11}
 & \multirow[c]{5}{*}{\rotatebox[origin=c]{90}{\textsc{Fair}}} & $\uparrow$ $\text{Jain}$ & 0.011 & 0.027 & 0.037 & \bfseries 0.115 & 0.003 & 0.006 & 0.024 & 0.069 \\
 &  & $\uparrow$ $\text{QF}^{*}$ & 0.044 & 0.068 & 0.145 & \bfseries 0.216 & 0.014 & 0.025 & 0.086 & 0.132 \\
 &  & $\uparrow$ $\text{Ent}$ & 0.407 & 0.514 & 0.596 & \bfseries 0.716 & 0.238 & 0.324 & 0.519 & 0.638 \\
 &  & $\uparrow$ $\text{FSat}^{*}$ & 0.044 & 0.068 & 0.145 & \bfseries 0.216 & 0.014 & 0.025 & 0.086 & 0.132 \\
 &  & $\downarrow$ $\text{Gini}$ & 0.987 & 0.971 & 0.945 & \bfseries 0.879 & 0.997 & 0.993 & 0.969 & 0.930 \\
\cline{2-11}
 & \multirow[c]{9}{*}{\rotatebox[origin=c]{90}{\textsc{Fair+Rel}}} & $\uparrow$ $\text{IBO}$ & 0.031 & 0.046 & 0.069 & \bfseries 0.091 & 0.012 & 0.018 & 0.054 & 0.074 \\
 &  & $\downarrow$ $\text{IWO}$ & 0.969 & 0.954 & 0.931 & \bfseries 0.909 & 0.988 & 0.982 & 0.946 & 0.926 \\
 &  & $\downarrow$ $\text{IAA}$ & 0.008 & 0.009 & 0.008 & 0.009 & 0.009 & 0.009 & \bfseries 0.008 & 0.009 \\
 &  & $\downarrow$ $\text{IFD}_{\div}$ & 0.018 & 0.012 & 0.019 & 0.011 & 0.016 & \bfseries 0.010 & 0.020 & 0.012 \\
 &  & $\downarrow$ $\text{IFD}_{\times}$ & 0.000 & 0.000 & 0.000 & \bfseries 0.000 & 0.000 & 0.000 & 0.000 & 0.000 \\
 &  & $\downarrow$ $\text{HD}$ & 0.221 & 0.255 & 0.226 & 0.262 & 0.265 & 0.273 & \bfseries 0.218 & 0.257 \\
 &  & $\downarrow$ $\text{MME}$ & 0.001 & 0.001 & 0.001 & \bfseries 0.001 & 0.003 & 0.001 & 0.001 & 0.001 \\
 &  & $\downarrow$ $\text{II-F}$ & 0.000 & 0.000 & 0.000 & 0.000 & 0.000 & 0.000 & \bfseries 0.000 & 0.000 \\
 &  & $\downarrow$ $\text{AI-F}$ & 0.000 & 0.000 & 0.000 & \bfseries 0.000 & 0.000 & 0.000 & 0.000 & 0.000 \\
\bottomrule
\end{tabular}
}
\end{minipage}
\raggedright
{\scriptsize *For ML-10M, QF $\equiv$ FSat, as QF is computed based on the 
\% of recommended items from all items, 
which in this case is equivalent to FSat. 
}
\end{table*}